# A Large Language Model Supported Synthesis of Contemporary Academic Integrity Research Trends

Thomas Lancaster

Imperial College London, United Kingdom

## Abstract

This paper reports on qualitative content analysis undertaken using ChatGPT, a Large Language Model (LLM), to identify primary research themes in current academic integrity research as well as the methodologies used to explore these areas. The analysis by the LLM identified 7 research themes and 13 key areas for exploration. The outcomes from the analysis suggest that much contemporary research in the academic integrity field is guided by technology. Technology is often explored as potential way of preventing academic misconduct, but this could also be a limiting factor when aiming to promote a culture of academic integrity. The findings underscore that LLM led research may be option in the academic integrity field, but that there is also a need for continued traditional research. The findings also indicate that researchers and educational providers should continue to develop policy and operational frameworks for academic integrity. This will help to ensure that academic standards are maintained across the wide range of settings that are present in modern education.

## Introduction

Academic integrity, a fundamental principle underpinning the credibility and quality of educational institutions worldwide, has garnered increasing attention in recent years. With the rapid evolution of educational technologies and methodologies, the landscape of academic integrity faces both new challenges and opportunities. This paper aims to provide an overview of the current research themes in academic integrity and the research approaches used to explore these themes, offering insights into the various aspects that shape this critical field. ChatGPT has been used as a guiding force in the construction of the paper, suggesting that a Large Language Model (LLM) like this may have a role to play in the future of academic integrity research and practice.

Academic integrity can be considered as an adherence to ethical and professional principles, standards, and values in education, research, and scholarship (ENAI, 2018). It embodies a commitment to fundamental values like honesty, trust, fairness, respect, responsibility, and courage (ICAI, 2021). These principles of behaviour allow academic communities to transform ideals into actions, guiding decisions and actions in the academic realm (Guerrero-Dib et al., 2020). Bertram Gallant (2016) emphasises that academic integrity should be integral to every academic endeavour, particularly in teaching and learning processes, aiming for academic excellence.

Examples of honest academic behaviours include being truthful, objective, crediting other authors' works, consistently applying policies, and addressing wrongdoings (ENAI, 2018; ICAI, 2021). In contrast, dishonest behaviours comprise acts like copying, using unauthorized materials during exams, plagiarism, collusion, falsifying data, impersonation, and various forms of contract cheating (Morris & Carroll, 2016). These behaviours are seen as fraudulent attempts to gain undeserved academic benefits and are often the focus of academic integrity research and assessments (McCabe, 2016; Rettinger & Bertram Gallant, 2022).

The concept of academic integrity should extend far beyond dishonest acts like these. Academic integrity encompasses a broad spectrum of ethical practices essential for the development of a reliable and trustworthy academic environment. Guerrero-Dib et al. (2023) give emphasis to the correlation between academic integrity and workplace ethics, advocating for demographic analysis, targeted interventions, continuous reminders of the benefits of ethical behaviour and the implementation of recognition schemes to incentivise it. They stress the importance of addressing academic dishonesty to prevent a culture of impunity and deterioration of ethical standards. Morris (2023) states that misconduct should be managed in a consistent, accessible and equitable manner, whilst recognising the diversity of the student body, and providing opportunities for discussion.

As these examples show, there is always a need to re-examine the traditional approaches used to uphold academic integrity, particularly with the proliferation of digital tools and the shift to blended and online learning. As such, understanding the current dynamics and dimensions of academic integrity is crucial for educators, administrators, and policymakers alike. Much of the current international work on academic integrity is disseminated as academic research publications, but such research also needs to be both inclusive and context sensitive.

This paper reports on a study that used ChatGPT to employ a qualitative content analysis methodology to analyse the prevalent research themes in academic integrity as of early 2023. Using the LLMs internal representations of research findings, which have been developed from sources including journal articles, conference proceedings, and educational reports, this paper identifies key focus areas in the field. The aim is to distil the essence of current research in academic integrity, providing a synthesised view that captures the nuanced views surrounding academic integrity and which can guide future research and practice.

## Background

Academic integrity is a both a practitioner field and a research field, although a possible criticism of the field is that research tends to evolve through practice rather than following a formalised process. The field appears strongly shaped by the development of technology, often due to the barriers to academic integrity that new technology can introduce. Levin (2012) emphasised the importance of rigor in action research on academic integrity, suggesting research partnering, controlling biases, standardized methods, and exploring alternative explanations to enhance the trustworthiness of research findings.

Several authors have previously attempted to capture the current state of academic research, although such publications can quickly become dated in a fast-moving field. For instance, Macfarlane et al. (2014) employed multivariate analysis to analyse academic integrity research as part of a comprehensive literature review and found the use of surveys and questionnaires to be commonplace. An example of the use of such an approach could be observed in Lanier (2006) who examined the prevalence of cheating in online versus traditional lecture courses, surveying 1,262 students.

An alternative method of exploring current research publications was introduced by Lancaster (2021), who used Natural Language Processing (NLP) techniques to analyse a dataset of 8,507 academic integrity papers. This approach, involving bigram analysis and sentiment analysis, provided insights into the linguistic patterns and sentiments in academic integrity research.

Documentary analysis is another prevalent method. Fielden and Joyce (2008) applied a multi-stakeholder, multi-level theoretical framework to analyse 125 published papers on academic integrity, focusing on the authors' moral judgments and views on academic integrity. Eaton et al. (2020) shared their methodological decisions regarding data collection in academic integrity policy analysis, emphasizing the importance of collaborative research teams and meticulous documentation.

Qualitative research methods, including interviews, seem to be integral to understanding the nuances of academic integrity. Macfarlane et al. (2014) incorporated interviews in their literature review, while Gaižauskaitė and Valavičienė (2023) highlighted the application of social sciences research methods, such as qualitative interviewing, in academic integrity research.

The increasing prevalence of online learning has also necessitated specific methodological approaches. Holden et al. (2021) provided a review of academic integrity in online assessment, addressing the motivations for cheating and methods for preventing and detecting dishonest behaviours in this context.

Overall, these varied methodologies reflect the complexity of academic integrity research, ranging from quantitative surveys to qualitative interviews and advanced NLP techniques. Each approach contributes uniquely to understanding the different facets of academic integrity. They also indicate why a current synthesis of academic integrity research will be so useful for providing future direction for the field.

**Methodology**

The focus of this paper is to identify current trends and themes of academic integrity research, as well as the methodological options for exploring those trends. Data to support this was gathered in January 2024 using ChatGPT Pro running GPT 4 and using ChatGPT's internal database of information. Fig. 1 shows the process followed to identify current academic integrity research themes. The process was developed and conducted by the ChatGPT LLM with human oversight.

Figure 1 – Methodological process overview flowchart

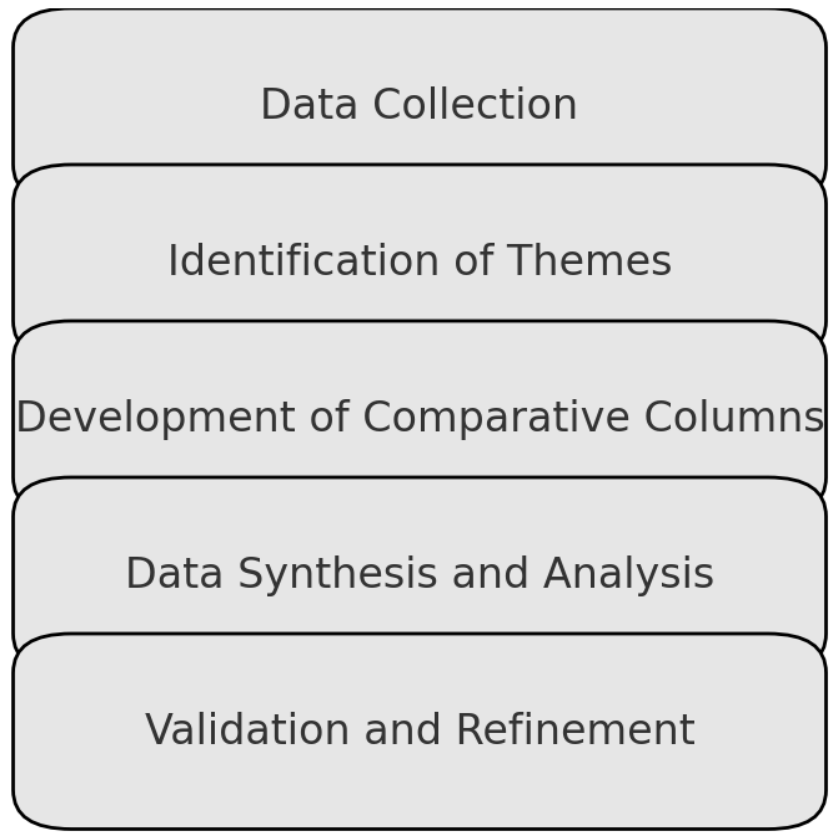

The methodology was implemented by ChatGPT, which reported details of its approach over the five stages as follows:

**Stage 1 - Data Collection**: ChatGPT accessed its internal database of information, which includes a wide range of academic literature, journal articles, conference proceedings, and educational reports up to April 2023. This database served as the primary source for extracting data relevant to the field of academic integrity.

**Stage 2 - Identification of Themes**: ChatGPT employed an iterative approach to identify major research themes within the academic integrity literature. Initial themes were discerned based on the recurrence and prominence of topics within the dataset. These themes were refined through comparative analysis to ensure accurate representation of key research areas.

**Stage 3 - Development of Comparative Columns**: For each identified theme, ChatGPT established several comparative columns, including 'Key Focus Areas', 'Methodologies Commonly Used', 'Recent Developments or Findings', and 'Geographic Focus'. This structure was chosen to enable a detailed and comparative overview of the research themes.

**Stage 4 - Data Synthesis and Analysis**: ChatGPT synthesized the data from its dataset according to the identified themes and comparative columns. This step involved organizing relevant information to reflect the current trends, methodologies, and geographic focuses within each theme.

**Stage 5 - Validation and Refinement**: The AI model reviewed the initial version of the table for accuracy, coherence, and comprehensiveness. Adjustments were made to ensure that the table accurately encapsulated the state of academic integrity research as of early 2023.

## Results

The structured and systematic overview generated of the academic integrity research landscape in shown in Table 1, which identified 7 themes, covering 13 key areas. The overview is based on data available to ChatGPT, consisting of sources up until April 2023.

| Research Theme | Key Focus Areas | Methodologies Commonly Used | Recent Developments or Findings | Geographic Focus |
|---|---|---|---|---|
| *Plagiarism Detection* | Software tools, Text-matching techniques | Computational analysis, Surveys | Advancements in AI-based detection systems | Global, with emphasis on Higher Education Institutions in North America and Europe |
| *Policy and Ethics in Education* | Development of integrity policies, Ethical education | Case studies, Policy analysis | Shift towards more holistic and preventative approaches | Predominantly North America and Europe |
| *Cheating Behaviours among Students* | Causes of cheating, Prevention strategies | Surveys, Psychological assessments | Increased focus on online learning environments | Global, with numerous studies in Asia and Australia |
| *Faculty Perspectives and Responses* | Training and awareness, Reporting behaviours | Interviews, Surveys | Emphasis on faculty training and support systems | North America, Europe, and some parts of Asia |
| *Cross-Cultural Comparisons* | Differences in perceptions and practices | Comparative studies, Ethnographic research | Insights into cultural influences on academic integrity | Global, with specific focus on comparisons between Western and Eastern educational systems |
| *Technology in Academic Integrity* | Use of online proctoring, Digital assessment tools | Technology assessment, User experience studies | Evaluation of the impact of remote proctoring and AI tools | Primarily in technologically advanced regions |
| *Legal and Institutional Frameworks* | Regulations impacting academic integrity, Institutional policies | Legal analysis, Institutional research | Analysis of legal responses to academic misconduct | Mostly in the United States, Europe, and some Asian countries |

Table 1 – The academic integrity research landscape

Fig. 2 shows the relationships between the research themes and the commonly used methodologies. It should be noted that this does not exclude methodologies being used to research themes that they are connected to in Fig. 2, but it does capture that surveys are often used as the de facto methodology for academic integrity research, even if other methodologies may be more appropriate.

Figure 2 – Relationship between research themes and methodologies

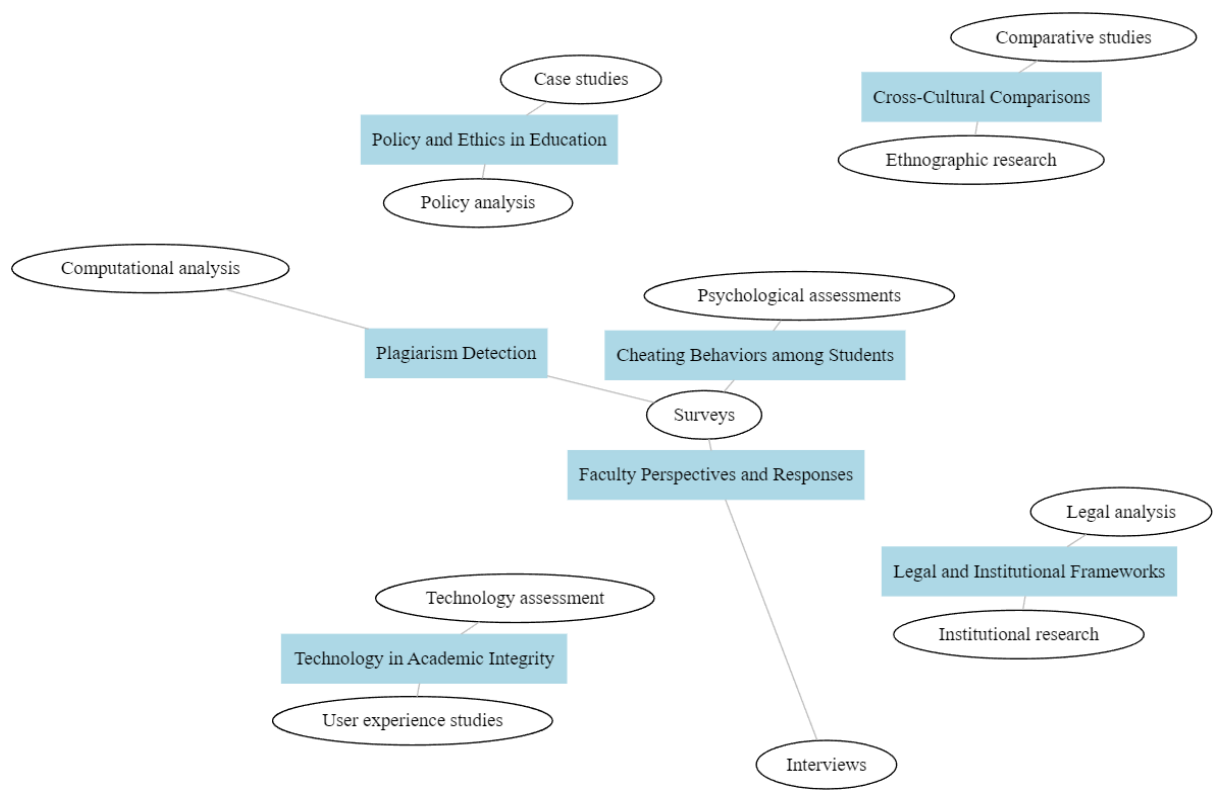

## Discussion and Conclusions

The findings of this study, as illustrated in Table 1, highlight the dynamic nature of academic integrity in the modern educational landscape, shaped by technological advancements and diverse cultural contexts. The study underscores the importance of shifting to a multifaceted approach to academic integrity, with current research addressing academic integrity from multiple viewpoints. This also indicates that most of the world is involved with academic integrity research, although the exact focus of that research varies by location.

One of the notable observations from the outcomes shown in Table 1 is the evolution of academic integrity issues with the advent of digital technologies. The proliferation of online resources and the ease of information access enabled by these have brought about new challenges in plagiarism and cheating. To counteract that, individuals and commercial companies alike have developed sophisticated detection tools and strategies. However, the ethical implications of surveillance tools like online proctoring software raise concerns about privacy and trust in educational environments and threaten to undermine the relationship between teaching staff and their students, where ICAI (2021) continue to recommend that trust is considered as a fundamental value of academic integrity.

The cultural dimensions of academic integrity revealed in this study underscore the need for context-specific strategies. The one-size-fits-all approach may be seen as inadequate. Educational institutions must consider cultural variances in understanding and practicing academic integrity to develop more effective and inclusive policies.

Faculty and institutional roles in upholding academic integrity continue to be a critical area. The findings suggest that while policy development is essential, the implementation and enforcement of these policies are equally important in creating a culture of integrity.

Emerging trends indicate a shift towards educative approaches to academic misconduct, moving away from purely punitive measures. This aligns with the recent focus on developing students' understanding and commitment to academic integrity, rather than solely penalising misconduct.

As the field of academic integrity continues to evolve, future research should focus on the long-term effectiveness of current strategies, the impact of emerging technologies, and the development of inclusive, culturally sensitive approaches. It is hoped that this paper will contribute to the ongoing dialogue in academic integrity, showing how LLMs like ChatGPT can be used in this field, whilst also providing insights and directions for educators, policymakers, and researchers in their efforts to ensure that academic integrity is at the heart of the educational system.

## References

Bertram Gallant, T. (2016). Systems approach to going forward. In T. Bretag (Ed.), *Handbook of academic integrity* (pp. 975–978). Springer. https://doi.org/10.1007/978-981-287-098-8_81

Bertram Gallant, T. (2017). Academic integrity as a teaching & learning issue: From theory to practice. *Theory Into Practice, 56*(2), 88-94. https://doi.org/10.1080/00405841.2017.1308173

Eaton, S. E., Stoesz, B. M., Thacker, E. J., & Miron, J. B. (2020). Methodological decisions in undertaking academic integrity policy analysis: Considerations for future research. *Canadian Perspectives on Academic Integrity, 3*(1), 83–91. https://doi.org/10.11575/cpai.v3i1.69768

ENAI. (2018). Glossary for academic integrity. https://academicintegrity.eu/wp/wp-content/uploads/2022/07/Glossary_revised_final.pdf

Fielden, K., & Joyce, D. (2008). An analysis of published research on academic integrity. *International Journal for Educational Integrity, 4*(2). https://doi.org/10.21913/IJEI.v4i2.411

Gaižauskaitė, I., & Valavičienė, N. (2023). Researching academic integrity: application of social sciences research methods. In *Academic Integrity in the Social Sciences: Perspectives on Pedagogy and Practice* (pp. 147-164). Cham: Springer International Publishing. https://doi.org/10.1007/978-3-031-43292-7_10

Guerrero-Dib, J.G., Portales, L., Gallego, D. (2023). Academic integrity as a way to promote workplace ethical behaviour. In: Curtis, G.J. (eds) Academic Integrity in the Social Sciences. Ethics and Integrity in Educational Contexts, vol 6. Springer, Cham. https://doi.org/10.1007/978-3-031-43292-7_11

Guerrero-Dib, J. G., Portales, L., & Heredia-Escorza, Y. (2020). Impact of academic integrity on workplace ethical behaviour. *International Journal for Educational Integrity, 16*(1). https://doi.org/10.1007/s40979-020-0051-3

Holden, O. L., Norris, M. E., & Kuhlmeier, V. A. (2021, July). Academic integrity in online assessment: A research review. In *Frontiers in Education* (Vol. 6, p. 639814). Frontiers Media SA. https://doi.org/10.3389/feduc.2021.639814

ICAI. (2021). *The fundamental values of academic integrity*. https://academicintegrity.org/resources/fundamental-values

Lancaster, T. (2021). Academic dishonesty or academic integrity? Using Natural Language Processing (NLP) techniques to investigate positive integrity in academic integrity research. *Journal of Academic Ethics, 19*(3), 363-383. https://doi.org/10.1007/s10805-021-09422-4

Lanier, M. M. (2006). Academic integrity and distance learning. *Journal of criminal justice education, 17*(2), 244-261. https://doi.org/10.1080/10511250600866166

Levin, M. (2012). Academic integrity in action research. *Action Research, 10*(2), 133-149. https://doi.org/10.1177/14767503124450

Macfarlane, B., Zhang, J., & Pun, A. (2014). Academic integrity: a review of the literature. *Studies in higher education, 39*(2), 339-358. https://doi.org/10.1080/03075079.2012.709495

McCabe, D. (2016). Cheating and honor: Lessons from a long-term research project. In T. Bretag (Ed.), *Handbook of academic integrity* (pp. 187–198). Springer. https://doi.org/10.1007/978-981-287-098-8_35

Morris, E.J. (2023). Integrating academic integrity: An educational approach. In: Eaton, S.E. (eds) Handbook of Academic Integrity. Springer, Singapore. https://doi.org/10.1007/978-981-287-079-7_96-1

Morris, E. J., & Carroll, J. (2016). Developing a sustainable holistic institutional approach: Dealing with realities "on the ground" when implementing an academic integrity policy. In T. Bretag (Ed.), *Handbook of academic integrity* (pp. 449–462). Springer. https://doi.org/10.1007/978-981-287-079-7_23-

Rettinger, D., & Bertram Gallant, T. (Eds.). (2022). Cheating academic integrity: Lessons from 30 years of research. Jossey-Bass.

## Acknowledgements

The author notes that ChatGPT was used to help develop this paper and to draft much of the text prior to human editing. This paper has itself been developed primarily as a case study for purposes of academic integrity and research integrity. It should be considered primarily as a case study of ChatGPT enabled research opportunities, rather than as an example of traditional peer reviewed research.